\documentclass[aps,twocolumn,nofootinbib,tightenlines,showpacs,%
showkeys,floatfix,tightenlines,preprintnumbers]{revtex4}

\usepackage{epsfig,bm}

\begin{document}



\title{The thermal width of heavy quarkonia moving in quark gluon plasma}


\author{Taesoo Song}%
\email{songtsoo@yonsei.ac.kr}

\author{Yongjae Park}%
\email{sfy@yonsei.ac.kr}

\author{Su Houng Lee}%
\email{suhoung@phya.yonsei.ac.kr}

\affiliation{Institute of Physics and Applied Physics,
Yonsei University, Seoul 120-749, Korea}

\author{Cheuk-Yin Wong}%
\email{cyw@ornl.gov}

\affiliation{Physics Division, Oak Ridge National Laboratory, Oak
Rige, TN 37830} \affiliation{Department of Physics, University of
Tennessee, Knoxville, TN 37996}


\begin{abstract}
The velocity dependence of the thermal width of heavy quarkonia
traveling with respect to the quark gluon plasma  is calculated up
to the NLO in perturbative QCD.   At the LO, the width decreases
with increasing speed, whereas at the NLO it increases with a
magnitude approximately proportional to the expectation value of the
relative velocity between the quarkonium and a parton in thermal
equilibrium. Such an asymptotic behavior is due to the NLO
dissociation cross section converging to a nonvanishing value in the
high energy limit.
\end{abstract}


\maketitle


\section{Introduction} Using arguments based on the contraction of the
Debye screening length in quark gluon plasma (QGP), Matsui and
Satz\cite{Matsui86} suggested $J/\psi$ suppression to be a signature
of the formation of QGP in the early stages of a heavy ion
collision. Indeed measurements at past SPS data \cite{Abreu:1997ji} showed nontrivial suppression patterns
that could be consistent with the original prediction.  However,
recent lattice calculations show that $J/\psi$ will survive past the
critical temperature $T_c$ for the phase transition
\cite{Hatsuda03,Hatsuda04,Datta03,Datta05,Datta06} up to about 1.6
$T_c$, while $\chi_c$ and $\psi'$ will dissolve above $T_c$.  These
findings suggest that a possible mechanisms for $J/\psi$
suppression could be the disappearance of feedbacks from the
$\chi_c$ and $\psi'$, together with the hadronic matter effects,
such as the  the nuclear absorption, the interactions with comovers,
and the shadowing effect \cite{Vogt:2005ia,Gunji:2007uy}.
Using a more precise determination of the cold nuclear matter absorption cross section of charmonium through p-A collisions \cite{Alessandro:2006jt}, it was found that such a feed down suppression scenario was indeed favored in results from semi-central Pb-Pb collisions \cite{Alessandro:2004ap} and from central In-In collisions at 158 GeV/nucleon \cite{Arnaldi:2006ee}.

Phenomenologically, the statistical hadronization model \cite{BraunMunzinger1,BraunMunzinger2,Becattini:1997ii,BraunMunzinger3,BraunMunzinger4} applied to the charmonium production \cite{BraunMunzinger2000} appears to be  compatible with RHIC data\cite{Andronic2007}.  In a kinetic model, the charmonium is produced in the whole temporal evolution of the QGP \cite{Thews,Grandchamp,Yan}. Further tests at LHC will discriminate between various pictures, and a unified picture is expected to emerge.

But before a simplified picture of $J/\psi$ suppression can be
adopted, detailed properties of $J/\psi$ above $T_c$ have to be
investigated.  Unfortunately, the present lattice calculations based
on Maximum Entropy Method still have poor resolution, and is not
able to reliably determine the thermal width or possible mass shift
above $T_c$ \cite{Hatsuda04,Datta03}.    However, there are several
recent works that can supplement the lattice calculation.   A recent
QCD sum rule calculation using the running coupling constant and the
gluon condensates extracted from a recent  lattice data shows that
the width of $J/\psi$ may be broadened or the mass reduced just over
$T_c$ \cite{Morita:2007pt}, which is consistent with results from
AdS/QCD \cite{Kim:2007rt}.   In another recent work, the
Debye screening length in the $J/\psi$ moving with respect to
QGP was calculated in AdS/QCD \cite{Liu:2006nn}.  In addition, the spectral functions of heavy quarkonia were extracted from the imaginary part of Green functions
obtained from potentials fitted to lattice data \cite{Cabrera:2006wh,Alberico:2006vw,Alberico:2007rg,Mocsy:2007yj}.
In another work, the thermal width of $J/\psi$ was investigated
using perturbative QCD up to the NLO \cite{Park}. The LO
perturbative QCD calculation for the heavy quarkonium dissociation
was invented by Peskin and Bhanot \cite{Peskin79,BP79} more than 20
years ago. Later, one of us redrived the LO result using
Bethe-Salpeter amplitude \cite{OKL02}, which was further used to
calculate the NLO results \cite{Song:2005yd}. In that work, the
bound state of quarkonium was described by Bethe-Salpeter amplitude,
and the perturbative QCD method was applied to calculate the decay
process. The binding energy of the quarkonium and its radius were
obtained by solving the Schr\"odinger equation with the potential
energy extracted from the lattice QCD calculation \cite{Wong04}. At
the NLO, it was found that the thermal width of $J/\psi$ increases
as temperature increases, while it decreases in the LO.   Moreover,
the total width was found to grow to more than 250 MeV at 1.4 $T_c$,
assuming the thermal quark gluon masses to be around  400 MeV.   If
the thermal width is so large, the $J/\psi$ initially formed at  1.6
$T_c$, will be dissociated immediately and not be able to escape the
quark gluon plasma until it cools down further to near $T_c$.

Another important aspect to be considered in a realistic heavy ion
collision is the velocity of the $J/\psi$ with respect to the QGP.
In particular, at LHC, more energetic heavy quarkonia are expected
to be produced and absorbed. Previously the Debye screening length
between a heavy quark and a heavy anti-quark pair moving with
respect to the QGP was calculated in a kinetic theory approach\cite{Chu:1988wh}.  It was found that the Debye screening length becomes shorter as the
velocity increases, because the parton density enhances in the heavy
quarkonium rest frame \cite{Chu:1988wh,Mustafa:2004hf}.   Recently,
the screening length was investigated in a AdS/CFT calculation
\cite{Liu:2006nn,Chernicoff:2006hi}, where it was found to be
approximately proportional to $[1-v^2]^{1/4}$, $v$ being the
velocity of $J/\psi$ with respect to the QGP. But whether AdS/CFT
calculations represents real QCD phenomena still remains
controversial. Therefore, in this work, we will extend a previous
NLO perturbative QCD calculation for the thermal width of a
quarkonium at rest \cite{Park} to that of a quarkonium moving with
respect to the QGP. The works mentioned above \cite{Chu:1988wh,Mustafa:2004hf,Liu:2006nn,Chernicoff:2006hi}
anticipated the shortening of the Debye screening length when the
heavy quark and anti-quark pair moves in the quark gluon matter or
equivalently when the matter is boosted, which may be called as the
change of static properties. On the other hand, our results provide
another component of the change through the thermal width of the
quarkonia as it moves through the  matter.  This may be called as
the change of dynamical properties.   The shortening of Debye
screening length means the radius and the binding energy of heavy
quarkonia should change as it moves through the matter.  However, in
this work, the radius and the binding energy of static quarkonia are
used for the purpose of investigating the change of a purely
dynamical property. We find that at the LO, the width decreases
with increasing $v$, which is caused by the vanishing of the
dissociation cross section of the quarkonium by an energetic gluon.
However, at the NLO the thermal width  increases with $v$. This is
due to the nonvanishing asymptotic cross section between the
quarkonium and an energetic parton at the NLO, which for the coulomb
wave function of the quarkonium scales as the square of the Bhor
radius. In section II, we briefly review formulas used throughout
this paper. In section III, we apply these formulas to $J/\psi$ and
$\Upsilon$. Some discussions are given in section IV. In the
Appendix, we derive the asymptotic form of the dissociation cross
section of a quarkonium in the high energy limit.

\section{Thermal width of a heavy quarkonium in pQCD}

The width of a hadron in the vacuum  comes from its spontaneous
decay. On the other hand, its  thermal width  results from its
interactions with the surrounding thermal particles. This thermal
width of a quarkonium moving with velocity $\beta$ in the medium is
defined as

\begin{eqnarray}
\Gamma^{eff}_\beta=d_p \int
\frac{d^3k}{(2\pi)^3}n(k_0)v_{rel}(\beta)\sigma(k,\beta),
\label{def-thermal-width}
\end{eqnarray}
where $d_p$ is the degeneracy factor, $n(k_0)$ is the distribution
function of the thermal particle, $v_{rel}$  the relative velocity
between decaying particle and the thermal particle, and
$\sigma(k,\beta)$ their energy-dependent elementary dissociation
cross section. Because we are interested in the decays of heavy
quarkonia in the quark-gluon plasma, the decaying particle is a
heavy quarkonium and the thermal particles are light quarks and
gluons. In this work, we considered only three light flavors. The
dissociation cross section can be written as
\begin{eqnarray}
\sigma(k,\beta)=\frac{1}{4v_{rel}(\beta) E_\Phi(\beta) E_p(k) }\int
d(p.s.) |\overline{M}|^2 \label{def-sigma}.
\end{eqnarray}

The first factor on the right side is the inverse of initial flux.
$E_\Phi(\beta)$, $E_p(k)$ are the energies of a quarkonium and a parton
respectively. $p.s.$ means phase space of final states, and
$\overline{M}$ is the spin-averaged invariant amplitude. The
invariant amplitudes for the decay of a quarkonium by partons are
listed in \cite{Park,Song:2005yd}. Substituting Eq.
(\ref{def-sigma}) into Eq. (\ref{def-thermal-width}) we find,

\begin{widetext}
\begin{eqnarray}
\Gamma^{eff}_\beta =d_p \int
\frac{d^3k}{(2\pi)^3}\frac{n(k \cdot u)}{4E_\Phi(\beta) E_p }\int d(p.s.)
|\overline{M}|^2 =\frac{d_p}{2 E_\Phi(\beta)} \int
\frac{d^4k}{(2\pi)^3}\frac{\delta^+(k^2-m_p^2)}{e^{ k\cdot u/T}\pm
1}\int d(p.s.) |\overline{M}|^2. \label{covariant}
\end{eqnarray}
\end{widetext}

Here the positive sign in the denominator is for a fermion and the
negative sign for a boson medium. The function $\delta^+$ means only
positive energy is allowed, and $u$ is the four velocity of the
thermal bath, hereafter called the Lab.\ frame.  In the Lab.\ fame,
$u=(1,0,0,0)$, whereas in the quarkonium rest frame,
$u=(1-\beta^2)^{-1/2}(1,0,0,-\beta)$. In the second equality of Eq.
(\ref{covariant}), all factors except $E_\Phi(\beta)$ are Lorentz
invariant. In fact, because $E_\Phi=\gamma m_\Phi$, where
$\gamma=1/\sqrt{1-\beta^2}$ and $m_\Phi$ is the mass of a
quarkonium, $\gamma \Gamma^{eff}_\beta$ is also Lorentz invariant
quantity, and we have

\begin{eqnarray}
\gamma \Gamma_\beta^{eff} = \gamma \Gamma_{Lab.}^{eff} =\Gamma_{\rm
quarkonium~rest~frame}^{eff},
\end{eqnarray}
where $\Gamma_{\rm quarkonium~rest~frame}^{eff}$ is calculated in
the rest frame of the quarkonium with a moving medium. This relation
just reflects time dilation.

\section{The thermal width of moving $J/\psi$ and $\Upsilon$ in QGP}

\begin{figure}
\centerline{
\includegraphics[width=4.5cm]{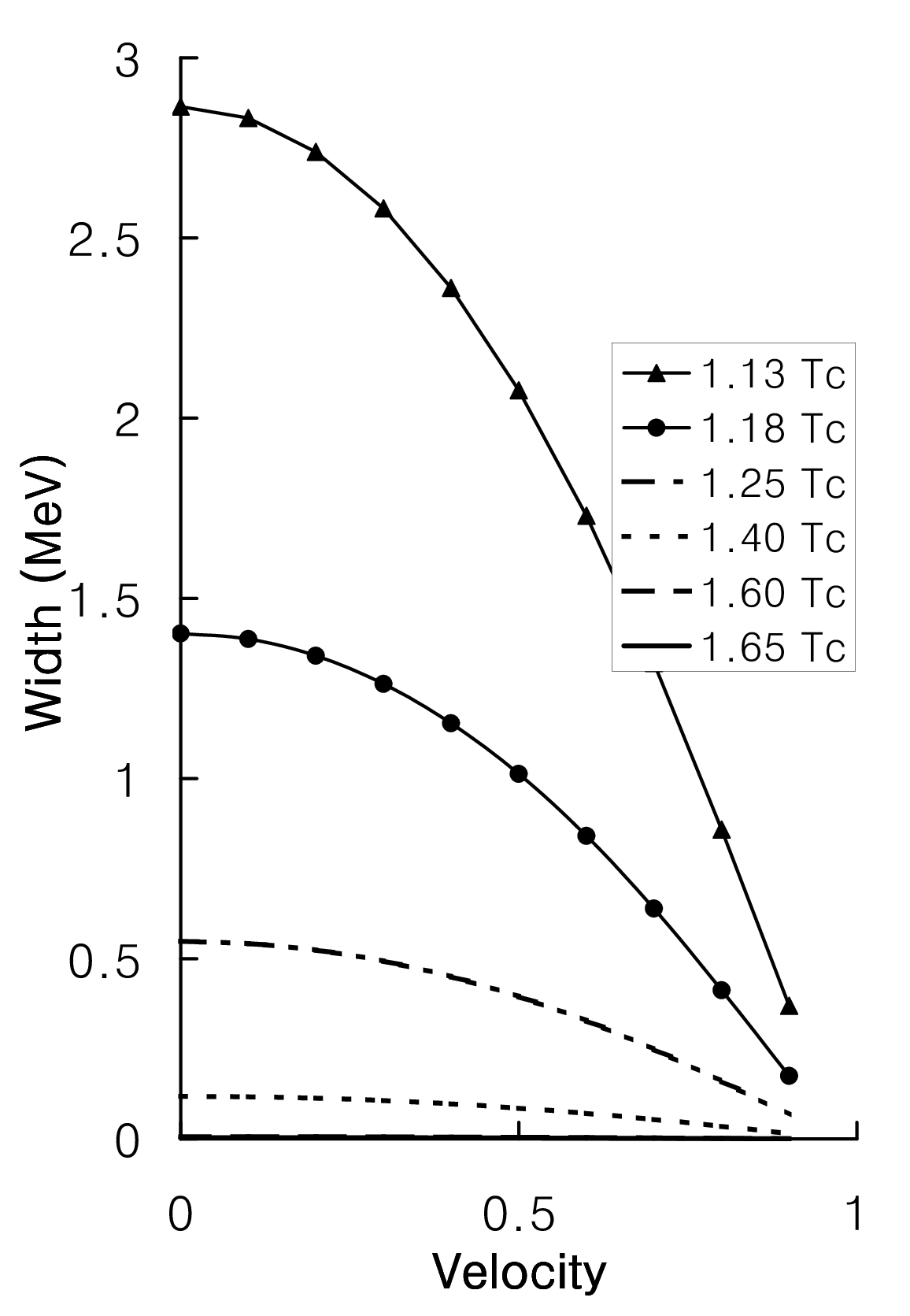}\hfill
\includegraphics[width=4.5cm]{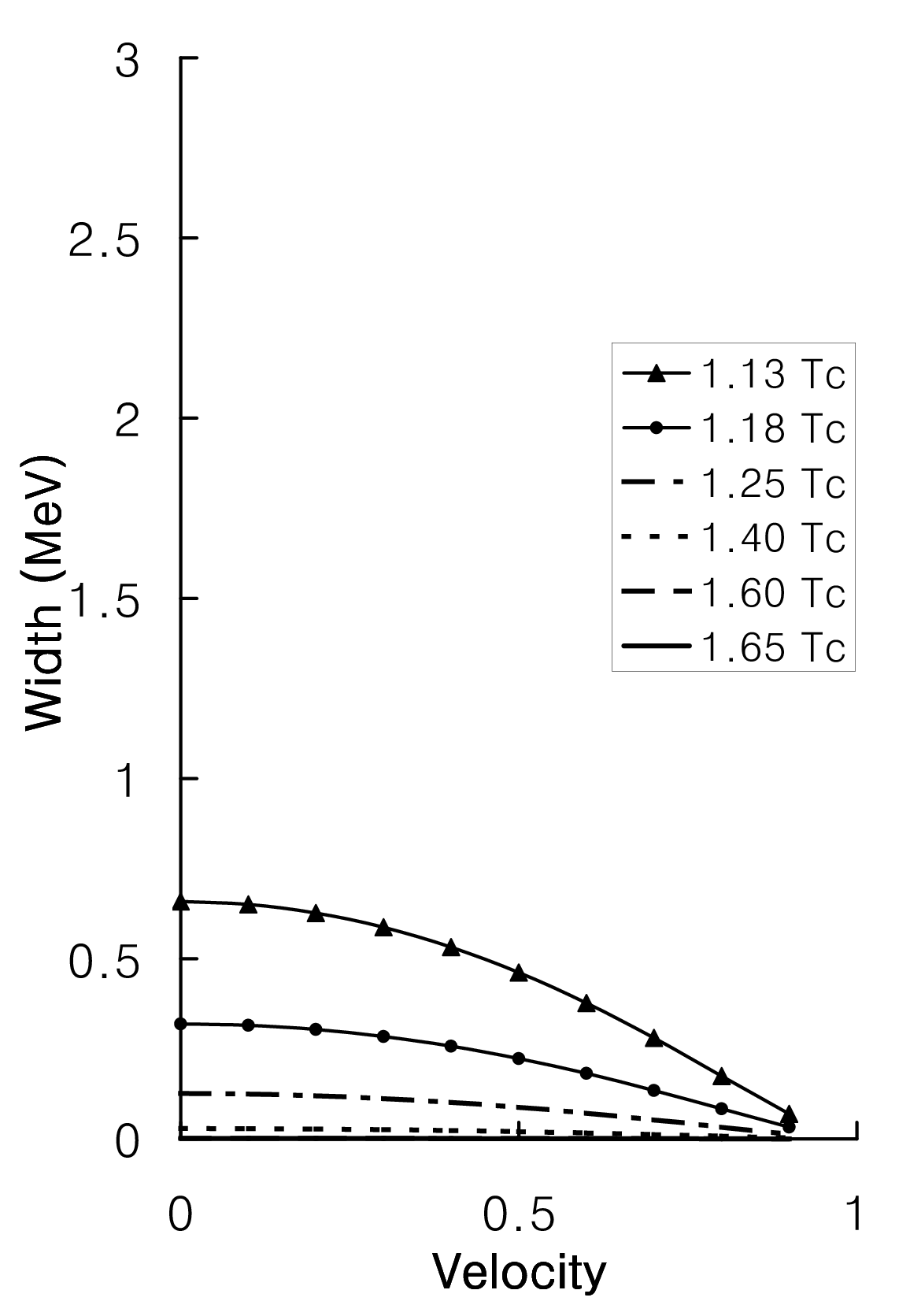}}
\centerline{
\includegraphics[width=4.5cm]{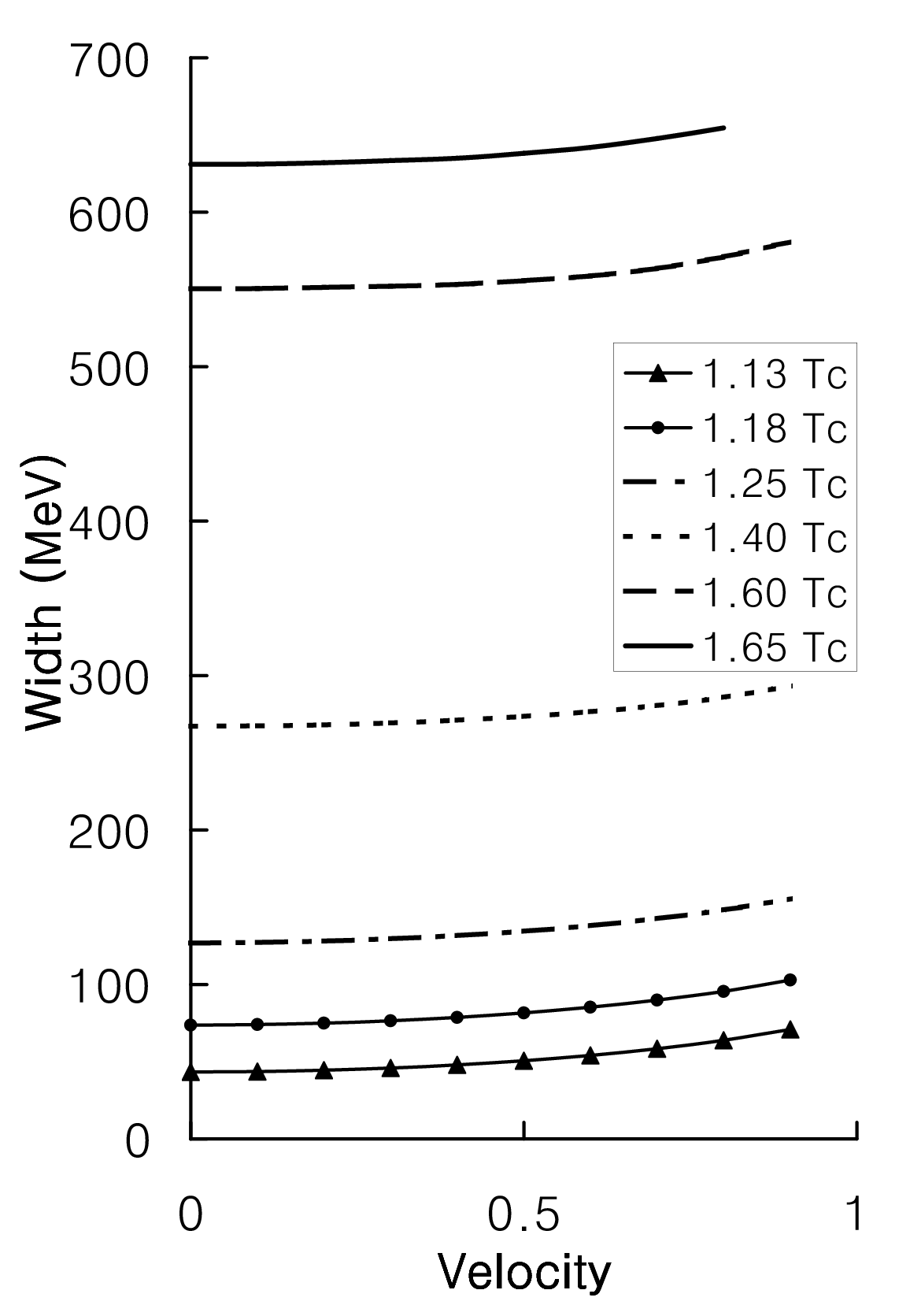}\hfill
\includegraphics[width=4.5cm]{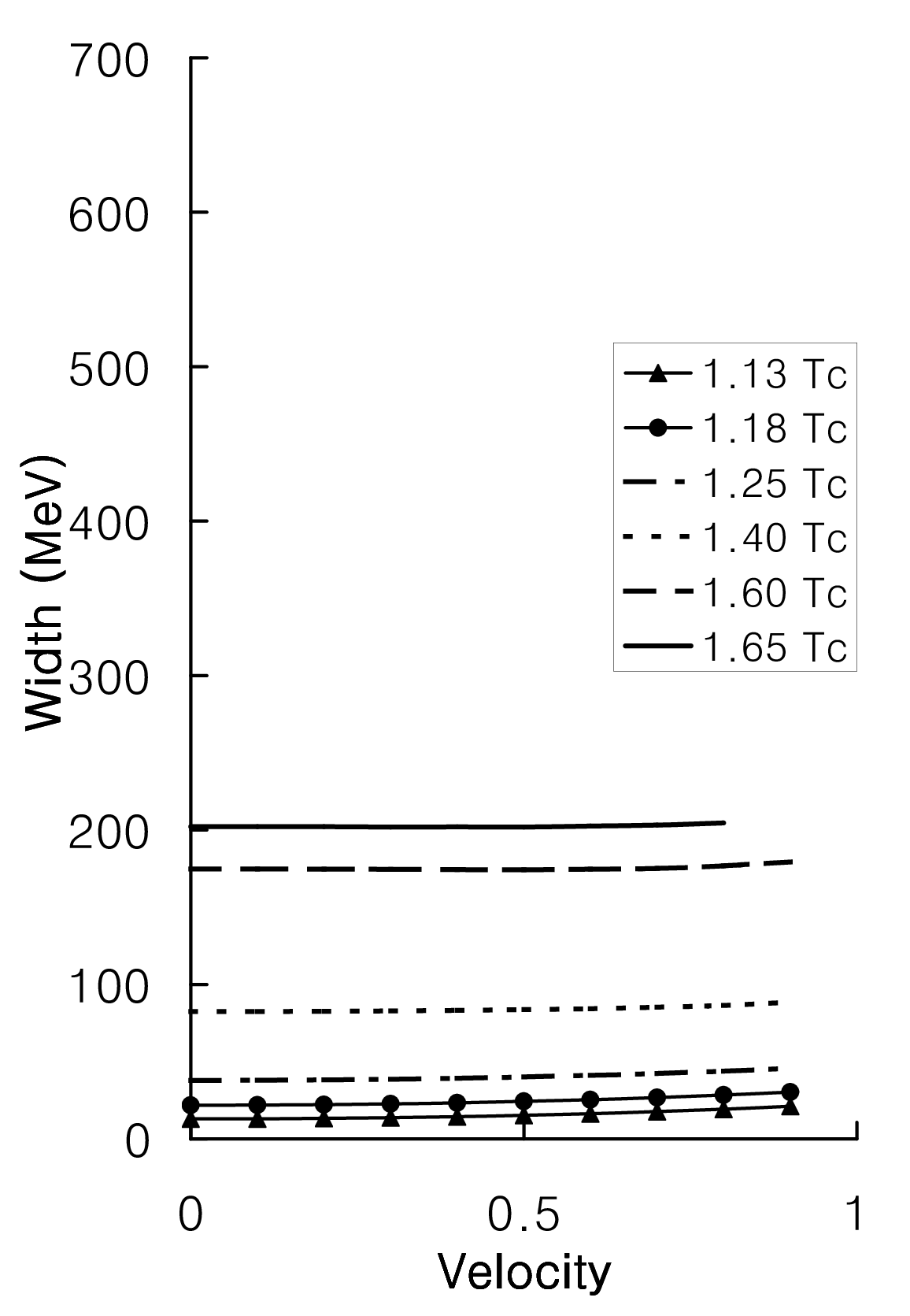}
}\caption{The variation of the thermal widths $\Gamma$ of $J/\psi$
in the LO(upper) and in the NLO(lower) as a function of their velocity in
QGP for various temperatures, and  assuming the thermal mass of a
parton to be 400 MeV(left) or 600 MeV(right) } \label{width-jpsi}
\end{figure}

We now apply the preceding formulas to calculate the thermal width
of the $J/\psi$  and the $\Upsilon$.   All calculation will be
performed in the Lab frame, and $\Gamma$ will be used to represent
$\Gamma^{eff}_{Lab.}$ from now on. The degeneracy factor $d_p$ in
Eq. (\ref{def-thermal-width}) is set to 16 for the gluons and to 36
for the quarks. For numerical purpose in these calculations, the
$T_c$ value is taken to be 170 MeV. The figures in the upper panel
of Fig. \ref{width-jpsi} show the variation of the thermal width
$\Gamma$ of $J/\psi$ in the LO of perturbative QCD as a function of
the velocity of the quarkonium in QGP. To the LO, the elementary
dissociation process is a thermal gluon dissociating the quarkonium
into a $\bar{c} c$ pair. Since the gluons acquire an effective
thermal mass in the QGP, we introduce a constant thermal mass of
either 400 or 600 MeV\cite{Levai}, respectively to represent a value
at the lower and upper limit.
 The graphs in the left
figure are obtained with a thermal gluon mass of 400 MeV, and the
right figures with 600 MeV.   The dissociation cross section in the
LO has the maximum value when the absorbed energy is slightly above
the threshold, and then rapidly decreases as the gluon energy
increases further \cite{OKL02,Song:2005yd}.  However, since the
thermal mass of an absorbed gluon is larger than the energy that
gives the maximum cross section, larger thermal mass will give
smaller cross section.  The binding energy of a $J/\psi$ in QGP is
obtained by solving the Schr\"{o}dinger equation with a potential
extracted from the lattice data, from which we find that it varies
from 36.4 MeV to few KeV as the temperature changes from 1.13 $T_c$
to 1.65 $T_c$ \cite{Park,Wong04}.  When the thermal mass of the
absorbed gluon becomes larger, the dissociation begins at higher
energy and the thermal width becomes smaller.  This is why the
thermal width with thermal gluon mass of  600 MeV is smaller than
that with 400 MeV. The graphs at the LO also show that the thermal
width decreases as the velocity of $J/\psi$ in QGP increases.   The
reason for it is again simple to understand.  As can be seen from
Eq. (\ref{def-thermal-width}), the thermal width is a convolution of
elementary cross section , the relative velocity between the
$J/\psi$ and the colliding parton, and the thermal distribution
function of the parton.
However, as $J/\psi$ moves across the parton matter, many slow
partons becomes fast partons as seen by the $J/\psi$.   As the
dissociation cross section by fast parton is small in the LO, due to
the small wave function overlap, the thermal width decreases.

The figures in the lower panel of Fig. \ref{width-jpsi} show the
thermal width of $J/\psi$ as a function of the velocity at the NLO,
when the thermal mass of a parton is 400 MeV(left) or 600 MeV(right)
respectively. These are the sum of quark induced and gluon induced
NLO processes.   As in the LO case, the width is smaller for larger
thermal parton mass.    This is mainly due to a larger virtuality of
the gluon propagator when the initial parton has a larger thermal
mass.    As shown in the figure, in contrast to the results at the
LO, the thermal widths slowly increase as the velocity of $J/\psi$
increases.   Such different behavior results from the asymptotic
form of the NLO dissociation cross section, which converges to some
finite value at the high energy limit.   In the appendix, we derive
the asymptotic form of the NLO dissociation cross section, assuming
the thermal parton mass to be sufficiently small.   If the cross
section is almost constant, the thermal width can be approximated as
follows,
\begin{eqnarray}
\Gamma \sim \sigma \int
\frac{d^3k}{(2\pi)^3}n(k_0)v_{rel}(\beta)\sim \sigma
<v_{rel}(\beta)>, \label{simf1}
\end{eqnarray}
where $<v_{rel}(\beta)>$ is the expectation value of the relative
velocity between  partons and the $J/\psi$ moving with velocity
$\beta$. As mentioned above, as $J/\psi$ moves faster, the
expectation value of the relative velocity increases. If the
$J/\psi$ moves with the speed of light, $<v_{rel}(\beta)>$
approaches 1, and $\Gamma \sim \sigma$. In conclusion, while  the
width decreases in the LO, its magnitude is small such that the sum
of the  LO and the NLO width increases with the velocity.

\begin{figure}
\centerline{
\includegraphics[width=4.5cm]{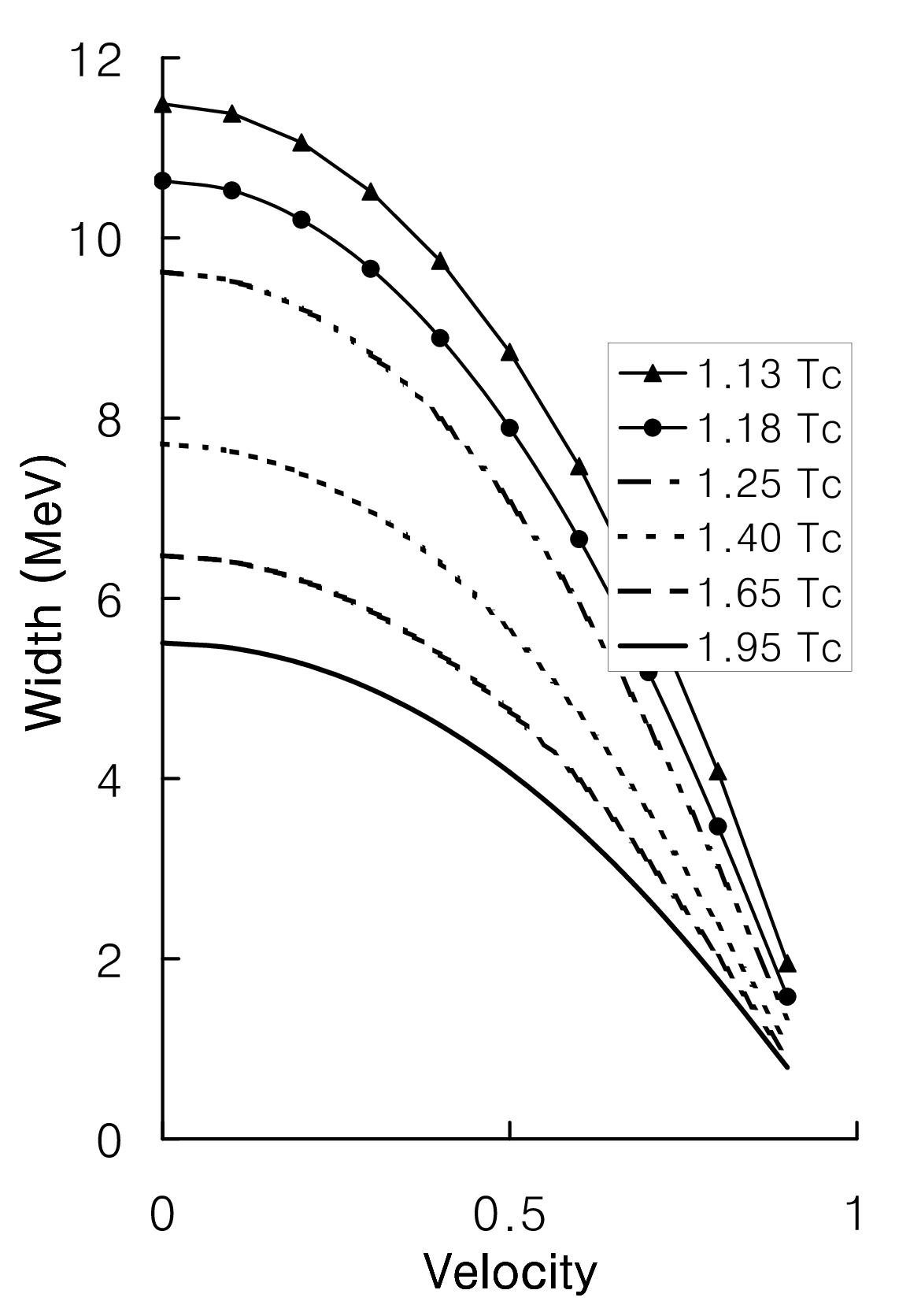}\hfill
\includegraphics[width=4.5cm]{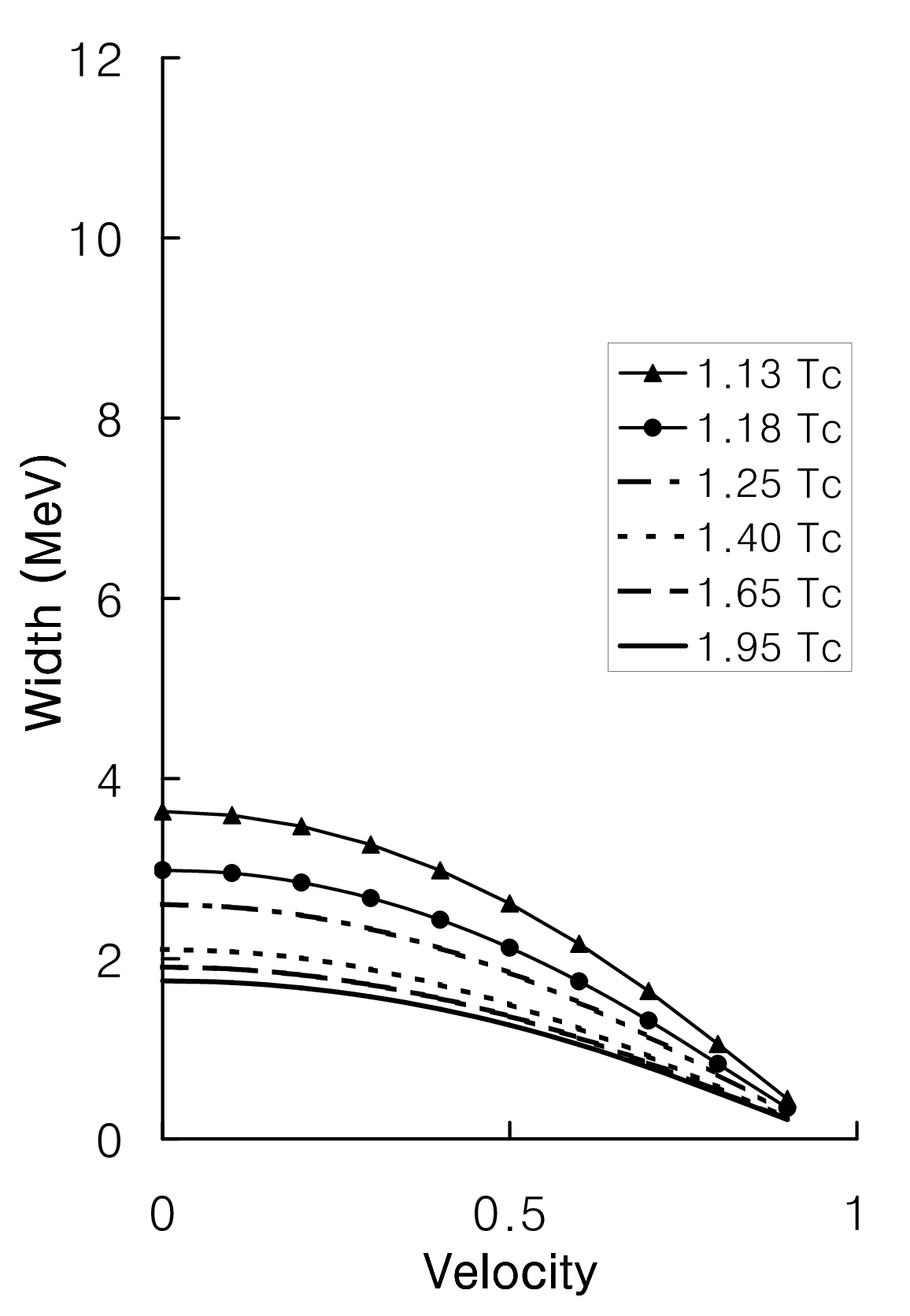}}
\centerline{
\includegraphics[width=4.5cm]{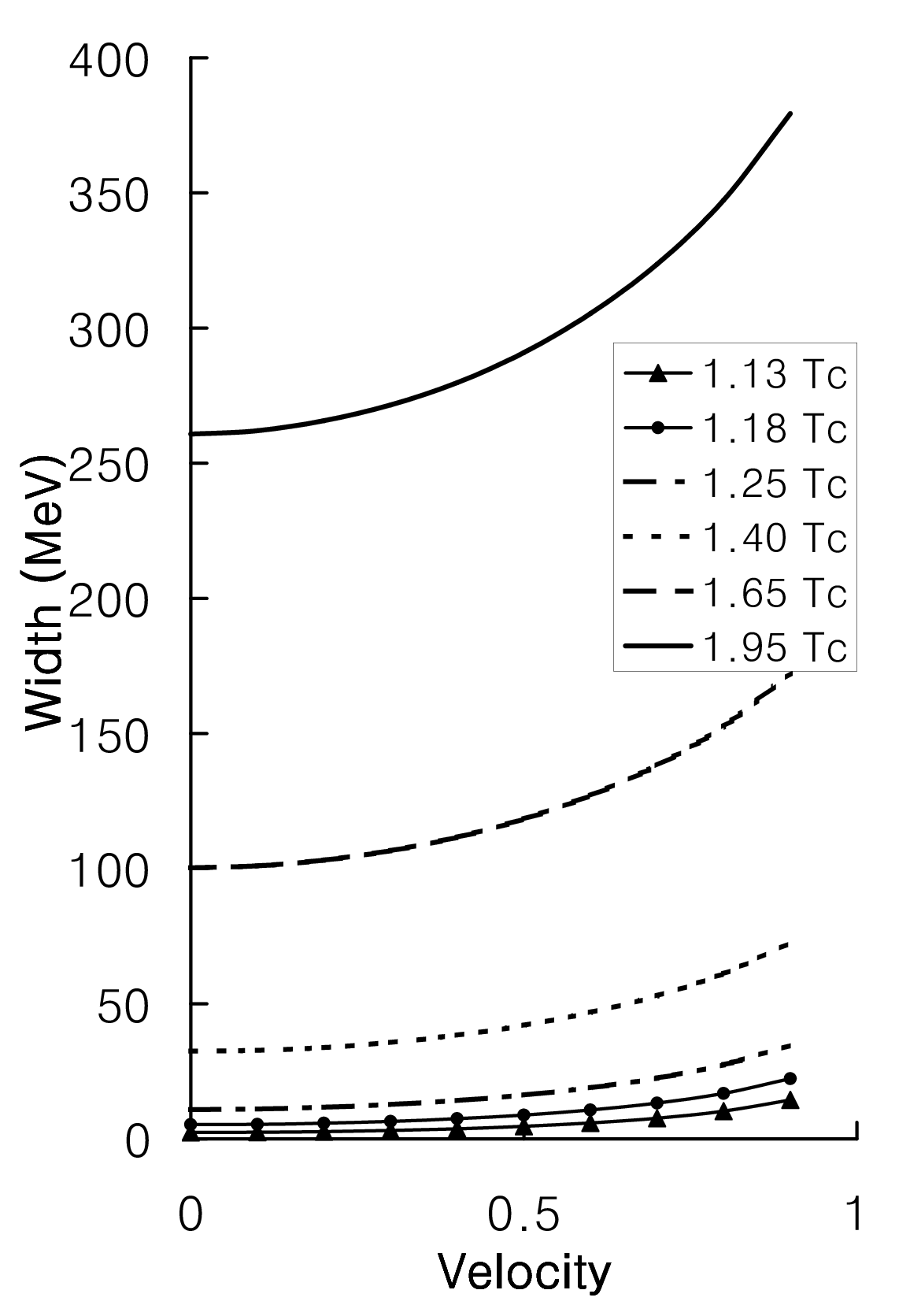}\hfill
\includegraphics[width=4.5cm]{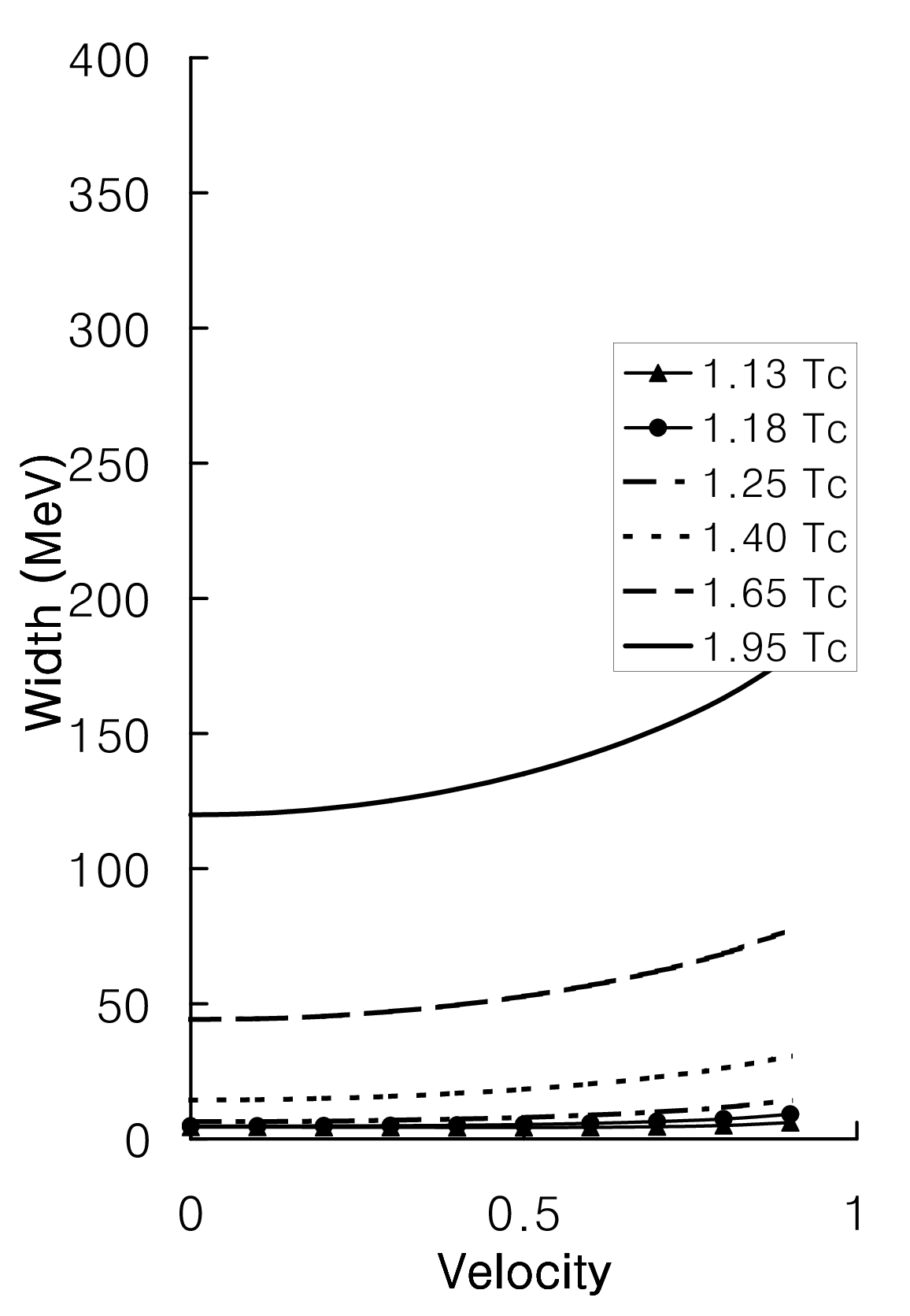}
}\caption{The variation of the thermal width $\Gamma$ of $\Upsilon$
in the LO(upper)and in the NLO(lower) as a function of their velocity with
the parton thermal mass of 400 MeV(left), of 600 MeV(right) }
\label{width-ups}
\end{figure}

Fig. \ref{width-ups} shows the thermal widths of $\Upsilon$ as a
function of its velocity in QGP.  As in the case of $J/\psi$, the
widths decrease in the LO, but increase in the NLO as the velocity
increases. However, the slope of increase is steeper for  $\Upsilon$
than for the $J/\psi$, which comes from the difference in their
binding energies.    If the binding energy is larger, the threshold
energy becomes larger and so does the energy at which the LO cross
section becomes maximum.  The dominant NLO process is a forward
scattering contribution, where the incoming parton emits a virtual
gluon, which then  dissociated the quarkonium via the LO process.
This causes the monotonically increasing NLO cross section to reach
its asymptotic value at a higher incoming energy when the binding
energy is larger.   This is the reason why for the $\Upsilon$
system, the thermal width has a larger slope when the velocity
increases.   The thermal width steadily increases until the
$\Upsilon$ is traveling close to the speed of light so that most of
the thermal partons have sufficient amount of energy to dissociate
the $\Upsilon$ at the asymptotic limit.   In contrast, for the
$J/\psi$ at rest, non trivial fraction of the thermal partons
already dissociates the quarkonium at the asymptotic limit.

\begin{figure}[b]
\centerline{
\includegraphics[width=4.5cm]{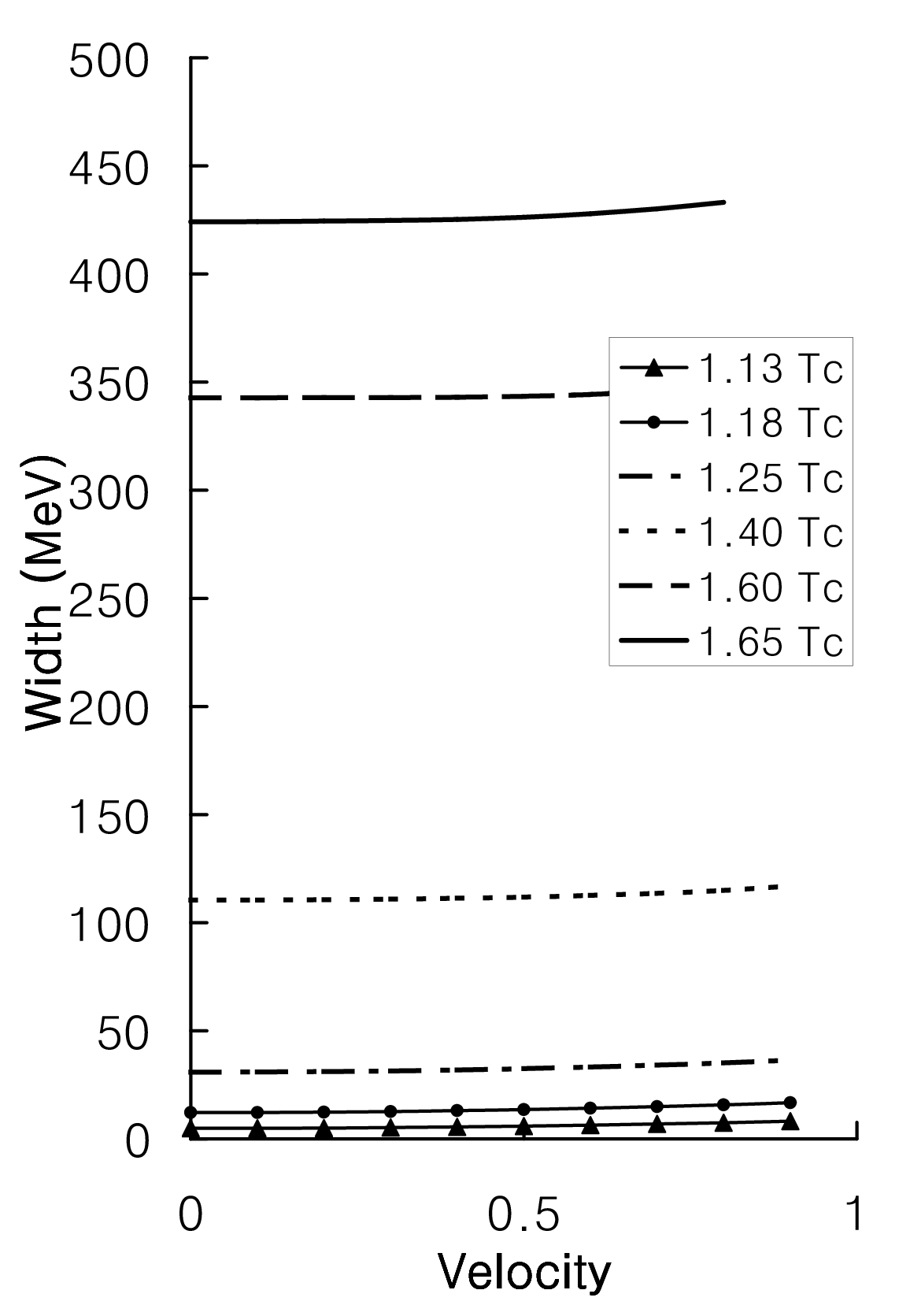}\hfill
\includegraphics[width=4.5cm]{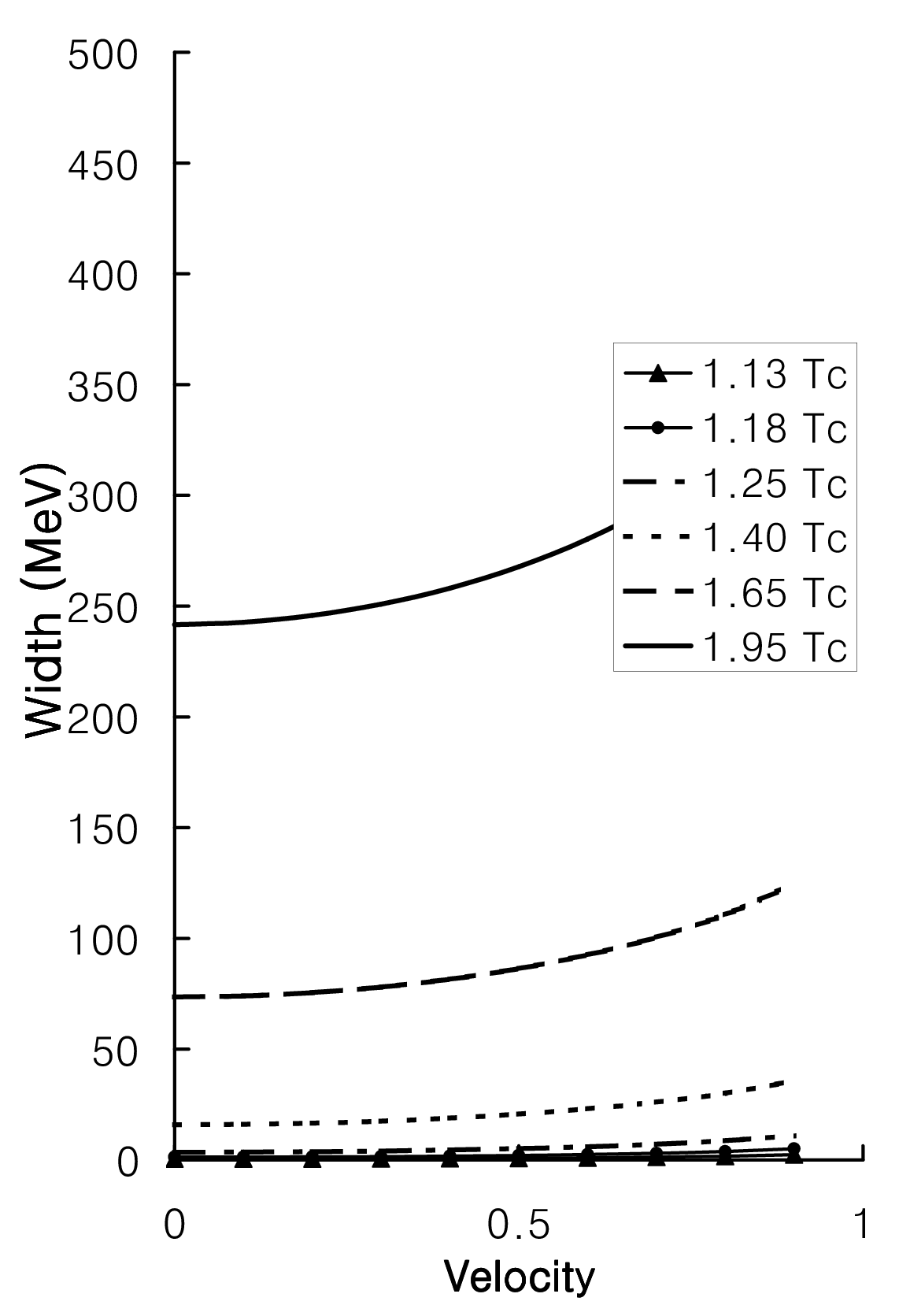}
}\caption{The variation of the thermal width $\Gamma$ of $J/\psi$ (left) and of $\Upsilon$ (right) from
the sum of LO and NLO as a function of their velocity, obtained with
the temperature dependent thermal mass \cite{Levai} }
\label{various-mass}
\end{figure}

Up to now, the thermal mass of partons were assumed to be temperature independent for simplicity.  However it is predicted to scale as $g(T)T$ by finite temperature QCD calculations.  Therefore, we finally present the result for the  thermal widths of $J/\psi$ and $\Upsilon$ obtained with temperature dependent thermal masses for the partons. The masses of thermal gluon and quark are taken respectively as,

\begin{eqnarray}
m_g^2(T)&=&\frac{g^2(T) T^2}{2}\bigg( \frac{N_c}{3}+\frac{N_F}{6} \bigg)\nonumber\\
m_q^2(T)&=& \frac{g^2(T) T^2}{3},
\end{eqnarray}
with
\begin{eqnarray}
g^2(T)=\frac{48\pi^2}{(11N_c-2N_f)\ln F^2(T,T_c,\Lambda)} \nonumber.
\end{eqnarray}
The number of color $N_c$ is set to 3, and the number of flavor $N_f$  to 3.  For the function  $F(T,T_c,\Lambda)$, we use the form obtained in \cite{Levai} from a fit to the lattice QCD calculations. Fig. \ref{various-mass} shows the thermal widths of $J/\psi$ and of $\Upsilon$ obtained with temperature dependent parton masses.  The magnitudes of the widths lie between the boundaries obtained with a constant thermal mass of 400 MeV and 600 MeV, as shown in Fig. \ref{width-jpsi} and in Fig. \ref{width-ups} for $J/\psi$ and $\Upsilon$ respectively.   This is so because the temperature dependent thermal mass lie between 400 MeV and 600 MeV in the  temperature range considered in this work.   Moreover, the functional dependencies  of the thermal widths with respect to the velocity are similar to those obtained with a constant thermal mass.

\section{Discussion}

In the heavy ion collisions at LHC, not only will the heavy
quarkonia be more amply produced, but be produced more
energetically. The suppression and/or relative enhancement of these
high $p_T$ quarkonia  will also occur in such an environment. In
this respect, it is very important to know how the properties of
heavy quarkonia will change when they move with respect to the QGP.
 Our result, based on the NLO perturbative QCD
calculation, shows that the thermal width becomes larger as the
quarkonium travels faster with respect to the QGP.   The rate of
increase is larger for the case of $\Upsilon$ case than for the
$J/\psi$, although the magnitude itself is larger for the latter.
Our result suggests that the survival rate of $J/\psi$ will be lower
than that obtained with the velocity independent thermal width
calculated at rest\cite{Park}.   Consequently, if  a relative
suppression of $J/\psi$ with higher than lower $p_T$ is observed, as
suggested by \cite{Liu:2006nn}, it may be a consequence of the
broadening of the thermal width of $J/\psi$ as well as of the
shortening of the Debye screening length.     The difference between
these two effects is that the former depends on the size of fireball
while the latter does not.  Therefore, a systematic study on the
A-dependence will be able to discriminate between these two effects.

\begin{acknowledgements} This work was supported by
the Korea Research Foundation KRF-2006-C00011.
\end{acknowledgements}

\hfil\break
\appendix
\centerline{\bf \large Appendix}
\bigskip Here, we derive the
asymptotic form of the cross section for the process $\Phi+q
\rightarrow Q+\bar{Q}+q$ in high energy limit. The cross section for
three-body decay is expressed as follow \cite{Park}

\begin{eqnarray}
\sigma&=&\frac{g^4m_Q^2 m_\Phi}{3\sqrt{(q\cdot k_1)^2-m_\Phi^2
m_{k_1}^2}}\int_\alpha^\beta dw^2 \frac{\sqrt{1-4m_Q^2/w^2}}{16^2
\pi^3 m_\Phi
|\vec{k_1}|}\nonumber\\
&\times& \int_{\alpha '}^{\beta '}dp_\Delta^2 \bigg|\frac{\partial
\psi({\bf p})}{\partial {\bf p }}\bigg|^2
\bigg(-\frac{1}{2}+\frac{k_{10}^2+k_{20}^2}{2k_1 \cdot k_2}\bigg),
\label{original}
\end{eqnarray}
where $q$, $k_1$, $k_2$ are the momentum of quarkonium $\Phi$,
incoming thermal quark, and outgoing thermal quark respectively, and
$p_\Delta^2=(k_1-k_2)^2, w^2=(q+p_\Delta)^2$. The momentum ${\bf p}$
is the relative three momentum between $Q$ and $\bar{Q}$. If the
quarkonium is the Coulomb bound state of $1S$, the absolute square
of derivative of quarkonium wavefunction is

\begin{eqnarray}
\bigg|\frac{\partial \psi({\bf p})}{\partial {\bf p
}}\bigg|^2&=&2^{10}\pi a_0^5\frac{a_0^2 {\bf p}^2}{(|a_0{\bf
p}|^2+1)^6}\nonumber\\
&=& 2^{10}\pi (a_0
\epsilon_0)^5\frac{k_{10}-k_{20}-\epsilon_0}{(k_{10}-k_{20})^6},
\end{eqnarray}
where $a_0$ is the Bohr radius and $\epsilon_0$ is the binding
energy of the quarkonium. In the second line of the above equation,
the energy conservation condition $ m_\Phi+k_{10}={\bf
p}^2/m_Q+k_{20}$, and the relation $a_0^2=1/(\epsilon_0 m_Q)$ have
been used.  It might seem odd to use the nonrelativistic energy
conservation condition in the high energy limit.  However, in the
high energy limit, the forward scattering is dominant, and
therefore, while the incoming quark energy is very large,  the
transferred energy carried by the virtual gluon that dominantly
dissociates the quarkonium  is not large. This is so because the LO
dissociation cross section is dominant near threshold.   Hence a
nonrelativistic treatment of the bound state kinematics is
justified. $\alpha$, $\beta$, $\alpha'$, and $\beta'$ are the
integration limits for drawing Dalitz plot on the plane of
$p_\Delta^2$ and $w^2$. They are respectively

\begin{eqnarray}
\alpha&=&4m_Q^2 \nonumber\\
\beta &=&(\sqrt{s}-m_{k_1})^2 \nonumber\\
\alpha '&=&-b-\sqrt{b^2-ac} \nonumber\\
\beta '&=&-b+\sqrt{b^2-ac},
\end{eqnarray}
where
\begin{eqnarray}
b&=&\{s-(m_\Phi+m_{k_1})^2\}\{s-(m_\Phi-m_{k_1})^2\}/(2s)\nonumber\\
&&-\{s-(m_\Phi^2-m_{k_1}^2)\}(w^2-m_\Phi^2)/(2s)\nonumber\\
b^2-ac&=&\{(s-m_\Phi^2+m_{k_1}^2)^2-4s m_{k_1}^2\}\nonumber\\
&&\times
 \{w^2-(u+m_{k_1})^2\}\{w^2-(u-m_{k_1})^2\}/(4s^2).\nonumber
\end{eqnarray}

Here, $s$ is the square of the initial energy in center-of-mass
frame. In the limit of $s \rightarrow \infty$,

\begin{eqnarray}
\beta  &\rightarrow& s \nonumber\\
\alpha ' &\rightarrow& w^2-s \nonumber\\
\beta ' &\rightarrow& 0 \nonumber\\
k_{10}, ~ |\vec{k_1}| &\rightarrow& \frac{s}{2m_\Phi}\nonumber\\
k_{20}&\rightarrow& \frac{s-w^2+p_\Delta^2}{2m_\Phi}\nonumber
\end{eqnarray}

and the initial flux becomes
\begin{eqnarray}
4\sqrt{(q\cdot k_1)^2-m_\Phi ^2 m_{k_1}^2}\rightarrow 2s.
\end{eqnarray}

Then the elementary cross section becomes

\begin{eqnarray}
\sigma&\approx&\frac{16 g^4 m_Q^2 m_\Phi (a_0 \epsilon_0)^5}{3 \pi^2
s^2}\int_{4m_Q^2}^{s}dw^2 \sqrt{1-\frac{4m_Q^2}{w^2}}\int_{w^2-s}^0
dp_\Delta^2
\nonumber\\
&&\times \frac{(k_{10}-k_{20}-\epsilon_0)}{(k_{10}-k_{20})^6}
\bigg[-\frac{1}{2}+\frac{(k_{10}-k_{20})^2}{2k_1\cdot
k_2}+\frac{k_{10}k_{20}}{k_1\cdot
k_2}\bigg]\nonumber\\
&\approx& \frac{16 g^4 m_Q^2 m_\Phi (a_0 \epsilon_0)^5}{3 \pi^2
s^2}\int_{4m_Q^2}^{s}dw^2 \sqrt{1-\frac{4m_Q^2}{w^2}}\int_{w^2-s}^0
dp_\Delta^2
\nonumber\\
&&\times \frac{(k_{10}-k_{20}-\epsilon_0)}{(k_{10}-k_{20})^6}
\frac{k_{10} k_{20}}{k_1\cdot k_2}. \label{bracket}
\end{eqnarray}

Because $k_{10}-k_{20}$ varies from $\epsilon_0$ approximately to
$s^{1/2}$, first two terms in the square bracket have no
contribution in the large $s$ limit. After integration with respect to
$p_\Delta^2$, the cross section becomes
\begin{eqnarray}
\sigma &\approx& \frac{2^8 g^4 m_Q^2 m_\Phi^4(a_0 \epsilon_0)^5}{3
\pi^2 }\int_{4m_Q^2}^{s}dw^2
\nonumber\\
&&\times \frac{\sqrt{1-4m_Q^2/w^2}}{(w^2-m_\Phi^2)^5}\bigg[
-\frac{25}{12}+\ln
\frac{w^2-m_\Phi^2}{2m_{k_1}^2}\nonumber\\
&& ~ ~ -\frac{2m_\Phi
\epsilon_0}{w^2-m_\Phi^2}\bigg(-\frac{137}{60}+\ln
\frac{w^2-m_\Phi^2}{2m_{k_1}^2}\bigg)\bigg], \label{first-int}
\end{eqnarray}
where we used the following decomposition formula for integration
with respect to $p_\Delta^2$,
\begin{eqnarray}
\frac{1}{(x-a)(x-b)^n}&=& \frac{-1}{a-b}\frac{1}{(x-b)^n}+\frac{-1}{(a-b)^2}\frac{1}{(x-b)^{n-1}}\nonumber\\
&...&
+\frac{-1}{(a-b)^n}\frac{1}{x-b}+\frac{1}{(a-b)^n}\frac{1}{x-a}\nonumber
\end{eqnarray}
and we ignored the thermal mass $m_{k_1}$ for simplicity except in
the logarithm as a regulator.  This elimination brings about a
potential problem when the binding energy is very small. To see its
subtlety, consider the decomposition formula in the following
example,
\begin{eqnarray}
\frac{1}{(2m_{k_1}^2-p_\Delta^2)(w^2-p_\Delta^2-m_{J/\psi}^2)}=
\frac{1}{w^2-m_\Phi^2-2m_{k_1}^2}\nonumber\\
\times \bigg\{
\frac{1}{2m_{k_1}^2-p_\Delta^2}-\frac{1}{w^2-p_\Delta^2-m_\Phi^2}\bigg\}.\nonumber\\
\label{separation}
\end{eqnarray}
Within the integration range, the left side of Eq.
(\ref{separation}) is always positive. However, if the binding
energy is very small, $1/(w^2-m_\Phi^2-2m_{k_1}^2)$ of the right
side of Eq. (\ref{separation}) has a singularity.  Of course,  the
term in the parenthesis on the right side diverges at the same time,
and the equality is maintained. However, if the thermal mass
$2m_{k_1}^2$ is dropped during the approximation for simplicity,
such divergence might not cancel and bring out the wrong result.

As the forward scattering becomes dominant in high energy
scattering, most of the  contributions come from the region $w^2\sim 4m_Q^2$
in the integration of $w^2$.  Therefore, the                                       the main suppression factor in the integrand in Eq.(\ref{first-int})
in the $s \rightarrow\infty$ limit is $1/(w^2-m_\Phi^2)^5$.  Terms such as
$\ln(w^2-m_\Phi^2)$ and $1/\sqrt{w^2}$ can  be replaced by
$\ln(4m_Q^2-m_\Phi^2)$ and $2m_Q$ respectively. Then we have,
\begin{eqnarray}
\sigma &\approx& \frac{2^7 g^4 m_Q m_\Psi^4(a_0 \epsilon_0)^5}{3
\pi^2 }\int_{4m_Q^2}^{s}dw^2
\frac{\sqrt{w^2-4m_Q^2}}{(w^2-m_\Phi^2)^5}\nonumber\\
&\times& \bigg[ -\frac{25}{12}+\ln
\frac{4m_Q^2-m_\Phi^2}{2m_{k_1}^2}\nonumber\\
&&~ -\frac{2m_\Phi
\epsilon_0}{w^2-m_\Phi^2}\bigg(-\frac{137}{60}+\ln
\frac{4m_Q^2-m_\Phi^2}{2m_{k_1}^2}\bigg)\bigg].
\end{eqnarray}

Using the beta function

\begin{eqnarray}
\int_0^\infty du \frac{u^m}{(1+u)^{m+n+2}}=\frac{m!~
n!}{(m+n+1)!}, \nonumber
\end{eqnarray}
the
$w^2$ dependent terms become

\begin{eqnarray}
\lim_{s \rightarrow \infty}\int_{4m_Q^2}^s dw^2
\frac{\sqrt{w^2-4m_Q^2}}{(w^2-m_\Phi^2)^5}=\frac{5\pi}{128(4m_Q^2-m_\Phi^2)^{7/2}} \nonumber \\
\lim_{s \rightarrow \infty} \int_{4m_Q^2}^s dw^2
\frac{\sqrt{w^2-4m_Q^2}}{(w^2-m_\Phi^2)^6}=\frac{7\pi}{256(4m_Q^2-m_\Phi^2)^{9/2}}, \nonumber
\end{eqnarray}
and finally the cross section becomes
\begin{eqnarray}
\sigma &\approx& \frac{2g^4 m_\Phi^4 a_o^2}{3\pi
m_Q^{1/2}(2m_Q+m_\Phi)^{9/2}} \bigg[ -\frac{125}{12}m_Q+\frac{167
}{60} m_\Phi\nonumber\\
&& ~ +(5m_Q-m_\Phi)\ln
\frac{\epsilon_o(2m_Q+m_\Phi)}{2m_{k_1}^2}\bigg]. \label{app1}
\end{eqnarray}

\begin{figure}
\centerline{
\includegraphics[width=6.5 cm, angle=270]{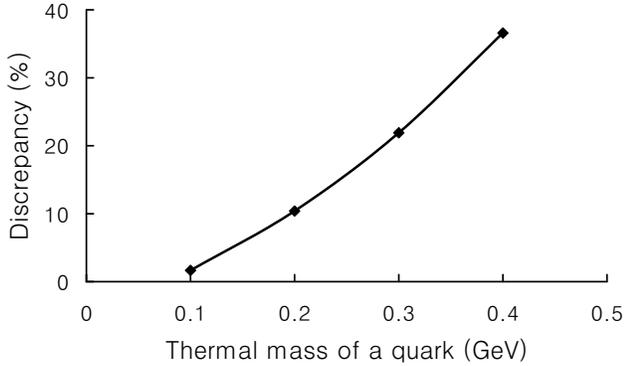}
}\caption{The discrepancy between Eq. (\ref{original}) and Eq.
(\ref{app1})} \label{discrepancy}
\end{figure}

Fig. \ref{discrepancy} shows the discrepancy between the result of
Eq. (\ref{original}) and the result of Eq. (\ref{app1}) at
$\sqrt{s}=$\ 100\ GeV, where we used the binding energy of $J/\psi$
in the vacuum \cite{Song:2005yd}. It is shown that the error
decreases as the thermal mass of a parton decreases. If we further
ignore the binding energy of a quarkonium, $m_\Phi=2m_Q$, we find
\begin{eqnarray}
\sigma \approx \frac{g^4 a_o^2}{48 \pi}\bigg( 3\ln\frac{2\epsilon_o
m_Q}{m_{k_1}^2}-\frac{97}{20}\bigg). \label{app2}
\end{eqnarray}
This formula has several important features.  First, the cross section is proportional to the square of the Bhor radius.  It is a consequence of the dipole type of cross section, which is proportional to the derivative of the momentum wave function squared.   The second important feature is the logarithmic term with argument proportional to the thermal mass, which acts as the regulator.

In the case of $\Phi+g \rightarrow Q+\bar{Q}+g$, the terms in the
square bracket of Eq. (\ref{bracket}) are replaced by the following
\cite{Song:2005yd,Park}

\begin{eqnarray}
&&\frac{k_1 \cdot k_2}{k_{10}k_{20}}
-4+\frac{2k_{10}}{k_{20}}+\frac{2k_{20}}{k_{10}}
-\frac{k_{20}^2}{k_{10}^2}-\frac{k_{10}^2}{k_{20}^2}\nonumber\\
&&+\frac{2}{k_1 \cdot k_2} \bigg[
(k_{10}-k_{20})^2\bigg\{\frac{(k_{10}+k_{20})^2}{k_{10}k_{20}}
-2\bigg\} +k_{10}k_{20}\bigg]. \nonumber\\
\label{nlog}
\end{eqnarray}
In the asymptotic limit, all but the last term in the square
bracket of Eq. (\ref{nlog}) are suppressed in the large $s$ limit.
Because its coefficient is twice of that for the quark induced case, the
asymptotic cross section is two times larger.  This ratio can also be seen in
the quark and gluon induced dissociation cross sections  shown in figures of ref \cite{Park}.

We remark that the asymptotic value for the cross section obtained
in the Appendix is for the coulomb bound state, while in the
previous sections,  both the Bohr radius $a_0$ and the binding
energy $\epsilon_0$ of quarkonia are extracted independently from
lattice data, such that the relation $a_0^2=1/(\epsilon_0 m_Q)$ is
not satisfied. Moreover, the extracted binding energy is too small
compared to the thermal mass of a parton, especially in the case of
$J/\psi$.   Because of these reasons, Eq. (\ref{app1}) or Eq.
(\ref{app2}) cannot be used directly.   Nevertheless, the property
that the dissociation cross section in the NLO converges to
nontrivial finite value in the high energy limit is still
maintained.


\begin{thebibliography}{10}

\bibitem{Matsui86}
  T.~Matsui and H.~Satz,
  Phys.\ Lett.\ B {\bf 178}, 416 (1986).


\bibitem{Abreu:1997ji}
  M.~C.~Abreu {\it et al.}  [NA50 Collaboration],
  Phys.\ Lett.\  B {\bf 410}, 327 (1997);
  M.~C.~Abreu {\it et al.}  [NA50 Collaboration],
  Phys.\ Lett.\  B {\bf 410}, 337 (1997);
  M.~C.~Abreu {\it et al.}  [NA50 Collaboration],
  Phys.\ Lett.\  B {\bf 450}, 456 (1999);
  M.~C.~Abreu {\it et al.}  [NA50 Collaboration],
  Phys.\ Lett.\  B {\bf 477}, 28 (2000).



\bibitem{Hatsuda03}
  M.~Asakawa, T.~Hatsuda and Y.~Nakahara,
  Prog.\ Part.\ Nucl.\ Phys.\  {\bf 46}, 459 (2001)
  [arXiv:hep-lat/0011040].


\bibitem{Hatsuda04}
  M.~Asakawa and T.~Hatsuda,
  Phys.\ Rev.\ Lett.\  {\bf 92}, 012001 (2004)
  [arXiv:hep-lat/0308034].


\bibitem{Datta03}
  S.~Datta, F.~Karsch, P.~Petreczky and I.~Wetzorke,
  Phys.\ Rev.\ D {\bf 69}, 094507 (2004)
  [arXiv:hep-lat/0312037].


\bibitem{Datta05}
  S.~Datta, F.~Karsch, P.~Petreczky and I.~Wetzorke,
  J.\ Phys.\ G {\bf 31}, S351 (2005)
  [arXiv:hep-lat/0412037].


\bibitem{Datta06}
  S.~Datta, A.~Jakovac, F.~Karsch and P.~Petreczky,
  AIP Conf.\ Proc.\  {\bf 842}, 35 (2006)
  [arXiv:hep-lat/0603002].


\bibitem{Vogt:2005ia}
  R.~Vogt,
  Acta Phys.\ Hung.\  A {\bf 25}, 97 (2006)
  [arXiv:nucl-th/0507027].


\bibitem{Gunji:2007uy}
  T.~Gunji, H.~Hamagaki, T.~Hatsuda and T.~Hirano,
  arXiv:hep-ph/0703061.

\bibitem{Alessandro:2006jt}
  B.~Alessandro {\it et al.}  [NA50 Collaboration],
  arXiv:nucl-ex/0612012.


\bibitem{Alessandro:2004ap}
  B.~Alessandro {\it et al.}  [NA50 Collaboration],
  Eur.\ Phys.\ J.\  C {\bf 39}, 335 (2005)
  [arXiv:hep-ex/0412036].


\bibitem{Arnaldi:2006ee}
  R.~Arnaldi {\it et al.}  [NA60 Collaboration],
  Nucl.\ Phys.\  A {\bf 774}, 711 (2006).



\bibitem{BraunMunzinger1}
  P.~Braun-Munzinger, J.~Stachel, J.~P.~Wessels and N.~Xu,
  Phys.\ Lett.\  B {\bf 344}, 43 (1995)
  [arXiv:nucl-th/9410026].

\bibitem{BraunMunzinger2}
  P.~Braun-Munzinger, J.~Stachel, J.~P.~Wessels and N.~Xu,
  Phys.\ Lett.\  B {\bf 365}, 1 (1996)
  [arXiv:nucl-th/9508020].

\bibitem{Becattini:1997ii}
  F.~Becattini, M.~Gazdzicki and J.~Sollfrank,
  Eur.\ Phys.\ J.\  C {\bf 5} (1998) 143
  [arXiv:hep-ph/9710529].


\bibitem{BraunMunzinger3}
  P.~Braun-Munzinger, I.~Heppe and J.~Stachel,
  Phys.\ Lett.\  B {\bf 465}, 15 (1999)
  [arXiv:nucl-th/9903010].

\bibitem{BraunMunzinger4}
  P.~Braun-Munzinger, D.~Magestro, K.~Redlich and J.~Stachel,
  Phys.\ Lett.\  B {\bf 518}, 41 (2001)
  [arXiv:hep-ph/0105229].

\bibitem{BraunMunzinger2000}
  P.~Braun-Munzinger and J.~Stachel,
  Phys.\ Lett.\  B {\bf 490}, 196 (2000)
  [arXiv:nucl-th/0007059].

\bibitem{The06}
R. L. Thews, Jour. Phys. G {\bf 32}, S401 (2006) and references
cited therein.

\bibitem{Andronic2007}
  A.~Andronic, P.~Braun-Munzinger, K.~Redlich and J.~Stachel,
  Phys.\ Lett.\  B {\bf 652}, 259 (2007)
  [arXiv:nucl-th/0701079].

\bibitem{Thews}
  R.~L.~Thews, M.~Schroedter and J.~Rafelski,
  Phys.\ Rev.\  C {\bf 63}, 054905 (2001)
  [arXiv:hep-ph/0007323].

\bibitem{Grandchamp}
  L.~Grandchamp, R.~Rapp and G.~E.~Brown,
  Phys.\ Rev.\ Lett.\  {\bf 92}, 212301 (2004)
  [arXiv:hep-ph/0306077].

\bibitem{Yan}
  L.~Yan, P.~Zhuang and N.~Xu,
  Phys.\ Rev.\ Lett.\  {\bf 97}, 232301 (2006)
  [arXiv:nucl-th/0608010].


\bibitem{Morita:2007pt}
  K.~Morita and S.~H.~Lee,
  arXiv:0704.2021 [nucl-th].


\bibitem{Kim:2007rt}
  Y.~Kim, J.~P.~Lee and S.~H.~Lee,
  Phys.\ Rev.\  D {\bf 75}, 114008 (2007)
  [arXiv:hep-ph/0703172].

\bibitem{Liu:2006nn}
  H.~Liu, K.~Rajagopal and U.~A.~Wiedemann,
  Phys.\ Rev.\ Lett.\  {\bf 98}, 182301 (2007)
  [arXiv:hep-ph/0607062].

\bibitem{Cabrera:2006wh}
  D.~Cabrera and R.~Rapp,
  arXiv:hep-ph/0611134.


\bibitem{Alberico:2006vw}
  W.~M.~Alberico, A.~Beraudo, A.~De Pace and A.~Molinari,
  Phys.\ Rev.\  D {\bf 75}, 074009 (2007)
  [arXiv:hep-ph/0612062].


\bibitem{Alberico:2007rg}
  W.~M.~Alberico, A.~Beraudo, A.~De Pace and A.~Molinari,
  arXiv:0706.2846 [hep-ph].


\bibitem{Mocsy:2007yj}
  A.~Mocsy and P.~Petreczky,
  arXiv:0705.2559 [hep-ph];
  A.~Mocsy and P.~Petreczky,
  arXiv:0706.2183 [hep-ph].


\bibitem{Park}
  Y.~Park, K.~I.~Kim, T.~Song, S.~H.~Lee and C.~Y.~Wong,
  arXiv:0704.3770 [hep-ph], to be published in Phys.\ Rev.\ C.


\bibitem{Peskin79}
M. E. Peskin, Nucl.\ Phys.\ B  {\bf 156} 365 (1979).


\bibitem{BP79}
G. Bhanot G and M. E. Peskin, Nucl.\ Phys.\ B  {\bf 156} 391 (1979).


\bibitem{OKL02}
  Y.~S.~Oh, S.~Kim and S.~H.~Lee,
  Phys.\ Rev.\ C {\bf 65}, 067901 (2002)
  [arXiv:hep-ph/0111132].


\bibitem{Song:2005yd}
  T.~Song and S.~H.~Lee,
  Phys.\ Rev.\  D {\bf 72}, 034002 (2005).


\bibitem{Wong04}
  C.~Y.~Wong,
  Phys.\ Rev.\ C {\bf 72}, 034906 (2005)
  [arXiv:hep-ph/0408020].


\bibitem{Chu:1988wh}
  M.~C.~Chu and T.~Matsui,
  Phys.\ Rev.\  D {\bf 39}, 1892 (1989).


\bibitem{Mustafa:2004hf}
  M.~G.~Mustafa, M.~H.~Thoma and P.~Chakraborty,
  Phys.\ Rev.\  C {\bf 71}, 017901 (2005).

\bibitem{Chernicoff:2006hi}
  M.~Chernicoff, J.~A.~Garcia and A.~Guijosa,
  JHEP {\bf 0609}, 068 (2006)
  [arXiv:hep-th/0607089].


\bibitem{Levai}
  P.~Levai and U.~W.~Heinz,
  Phys.\ Rev.\  C {\bf 57}, 1879 (1998)
  [arXiv:hep-ph/9710463].

\end{thebibliography}
\end{document}